# Unusual Bandgap Oscillations in Template-Directed π-Conjugated Porphyrin Nanotubes

Sarah I. Allec, Niranjan V. Ilawe, and Bryan M. Wong*

Department of Chemical & Environmental Engineering and Materials Science & Engineering Program

University of California-Riverside, Riverside, CA 92521, USA

*Corresponding author. E-mail: bryan.wong@ucr.edu. Web: http://www.bmwong-group.com

**Abstract**

Using large-scale DFT calculations (up to 1,476 atoms and 18,432 orbitals), we present the first detailed analysis on the unusual electronic properties of recently synthesized porphyrin nanotubes. We surprisingly observe *extremely* large oscillations in the bandgap of these nanostructures as a function of size, in contradiction to typical quantum confinement effects (i.e., the bandgap increases with size in several of these nanotubes). In particular, we find that these intriguing electronic oscillations arise from a size-dependent alternation of aromatic/non-aromatic characteristics in these porphyrin nanotubes. Our analyses of band structures and orbital diagrams indicate that the electronic transitions in these nanostructures are direct-bandgap, optically active "bright" states that can be readily observed in photoelectron spectroscopic experiments. Most importantly due to their unusual bandgap oscillations, we find that *both* type I and type II donor-acceptor p-n heterojunctions are possible in these template-directed, "bottom-up synthesized" porphyrin nanotubes – a unique property that is not present in conventional carbon nanotubes.

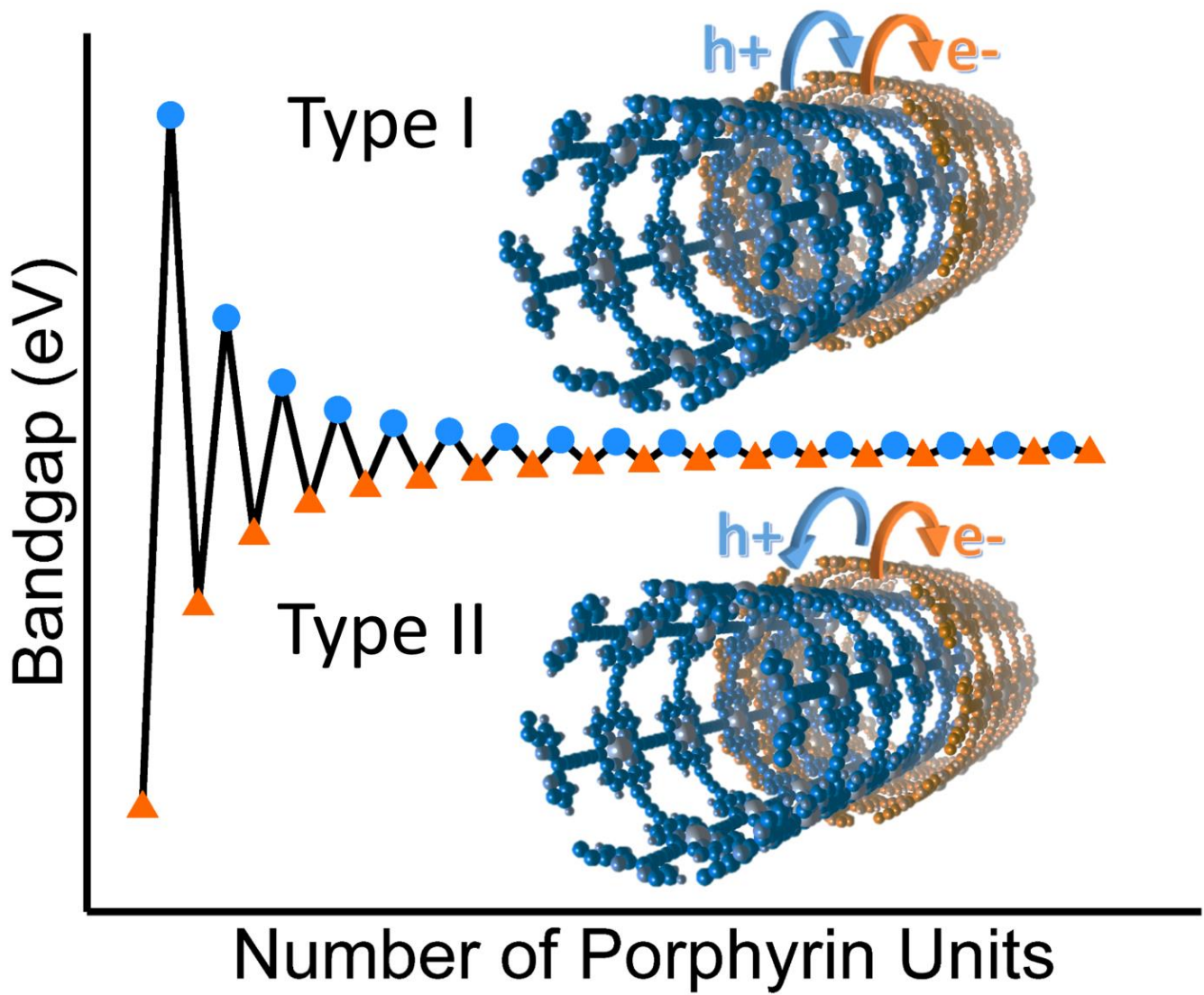

Table of Contents Graphic (large size)

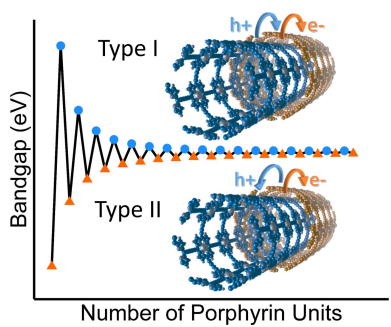

Table of Contents Graphic (small size)

Ever since Kekulé's seminal work on aromaticity,[1] nanostructures with π-conjugated orbitals have continued to provide fascinating systems for understanding and probing the unique electronic properties in nanomaterials. Although π-conjugation effects are intrinsically quantum-mechanical in nature (i.e., π-orbital delocalization/aromaticity has no classical analogue), these electronic effects have a profound impact on *real* material properties such as energy transfer, charge delocalization, and electron/hole mobility. Indeed, it is precisely these unique electronic effects that have ignited immense efforts in the synthesis of new conjugated nanostructures with the desired structural and electronic properties. While the electronic properties of carbon nanotubes (CNTs) can be tuned as a function of size, chirality, and allotrope type, the limitations of post-processing sorting techniques to obtain CNTs with identical properties has hampered their widespread use in technological applications. To circumvent these severe limitations, there have been recent efforts to obtain chirality-specific CNTs using a "growth-from-template" strategy for the bottom-up synthesis of these nanostructures. Specifically, in a series of joint theoretical-experimental investigations, we and the Jasti group have shown that cycloparaphenylenes form the fundamental annular segments of CNTs which can further serve as synthetic templates for armchair nanotubes.[2-6] These studies have, in turn, sparked additional research on the synthesis of other CNT chiralities[7-9] as well as other types of nanotubes built from nanoring molecular subunits.

Very recently in 2015, the Anderson group carried out the first experimental synthesis of a π-conjugated porphyrin nanotube based on a Venier template-directed synthesis of fused porphyrin nanorings.[10] Their approach was particularly noteworthy since the Vernier templating technique allows for the controlled synthesis of porphyrin nanorings of a specific size,[11] even up to sizes larger than many enzymes (diameters ~ 10 nm).[12] These rings have been stacked into supramolecular columns,[13] assembled into concentric ring structures[14], and recently resulted in a twelve-porphyrin nanotube that is fully conjugated in three dimensions.[10] These nanotubes can be viewed structurally as conjoined porphyrin rings, with a length longer than other synthesized π-conjugated molecular belts, barrels, or nanotubes[3, 15-20] (cf. Figure 1). Interestingly, in addition to their π-conjugated electronic structures,[21] these porphyrin nanostructures are strikingly similar to biological light-harvesting complexes.[22, 23] Because of the strong electronic coupling around the porphyrin nanoring, the absorption of light generates excited states that are delocalized over the entire circumference, acting as a unified π-system instead of distinct chromophores.[24]

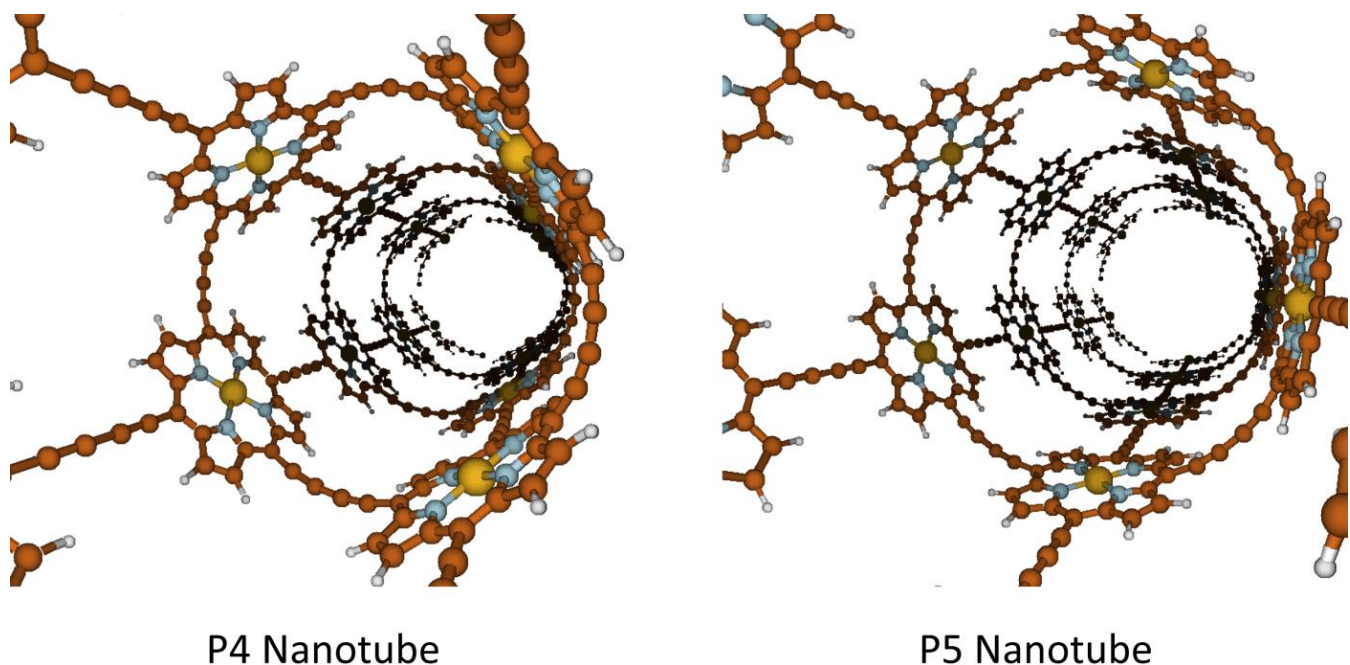

P4 Nanotube                              P5 Nanotube

**Figure 1.** Optimized structures of the P4 and P5 porphyrin nanotubes composed of 4 and 5 porphyrin subunits, respectively, around the nanotube circumference.

In this letter we present the first detailed analysis on the unusual structural and electronic properties of these novel porphyrin nanotubes. The various nanotube structures were created by linking together 2 to 36 individual porphyrin subunits (using acetylenic linkages), which form nanostructures that progressively span diameters ranging from 0.6 nm to 15.4 nm. Using large-scale density functional theory (DFT) calculations (up to 1,476 atoms and 18,432 basis functions), we surprisingly observe *extremely* large oscillations in the electronic bandgap of these nanostructures as a function of size, *in contradiction to typical quantum confinement effects* (i.e., the bandgap increases with size in several of these nanotubes). We find that these intriguing electronic oscillations arise from a size-dependent alternation of aromatic/non-aromatic characteristics in these porphyrin nanotubes – a unique property that is not present in conventional CNTs. We give a detailed analysis of these effects below and discuss the implications of these unusual properties that can be further leveraged for the controlled synthesis of new donor-acceptor and electronic materials.

All of the DFT calculations in this study were carried out with a massively-parallelized version of the CRYSTAL14 program in conjunction with the range-separated HSE06 functional and DZP basis set. We also explored other functionals and other DFT-based programs (including Gaussian and VASP) and found that the unusual bandgap oscillations in our study were not affected by our specific choice of functional or computational implementation. Further specifics of our calculations are given in the Computational Details section.

Figure 2 plots the electronic bandgap of the porphyrin nanotubes, which surprisingly shows large oscillations as a function of nanotube size/circumference. In particular, our calculations demonstrate that these effects are *not* small perturbations in the electronic structure, and these global oscillations span a wide energy range of up to 0.32 eV. It is also worth mentioning that graphene nanoribbons also exhibit bandgap oscillations as a function of size;[25] however, the oscillations in nanoribbons are due to a completely different phenomenon related to edge states and shape, whereas these edge effects are completely absent in the porphyrin nanostructures studied here. While our calculations focus on porphyrin nanotubes (which are periodic along the nanotube axis), we also observe similar oscillations in the individual porphyrin molecular nanorings themselves. Moreover, we note that porphyrin nanotubes composed of an even number of subunits have bandgaps that increase with size (in contradiction with typical quantum confinement effects) whereas nanotubes with an odd number of subunits have bandgaps that decrease with size. As the number of porphyrin units increases, a limiting value of 0.81 eV is obtained, which recovers the bandgap of the "parent" planar porphyrin sheet (see Supporting Information).

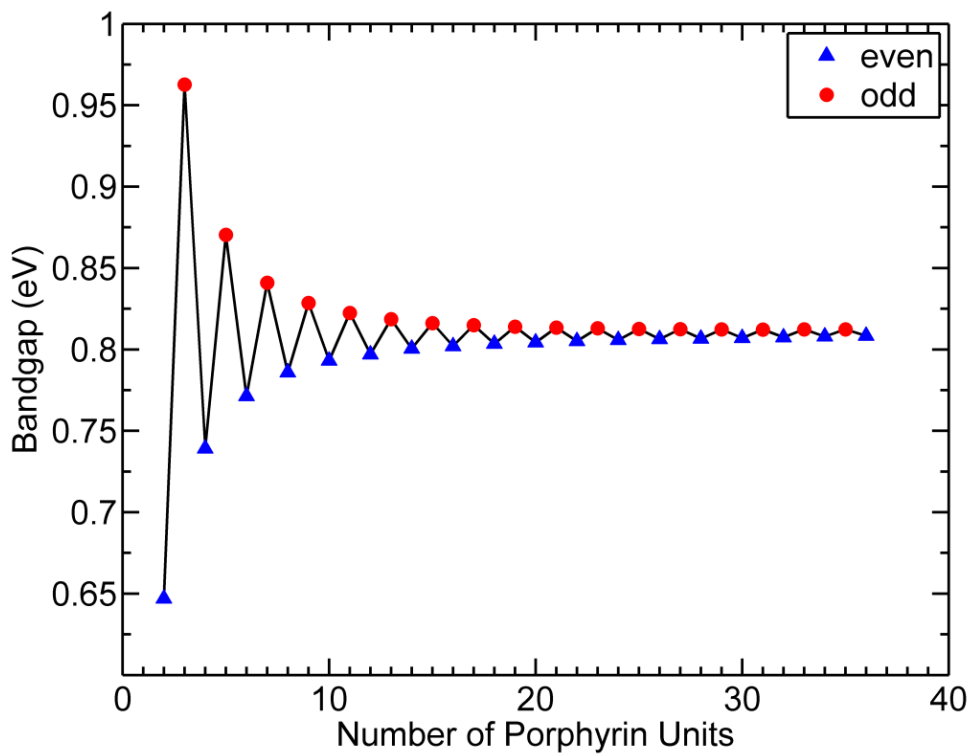

**Figure 2.** Electronic bandgaps for various porphyrin nanotubes as a function of the number of porphyrin units obtained at the HSE06/DZP level of theory. Porphyrin nanotubes composed of an even number of subunits have bandgaps that increase with size (in contradiction with typical quantum confinement effects) whereas nanotubes with an odd number of subunits have bandgaps that decrease with size.

Two questions now naturally arise: (1) What is the origin of these bandgap oscillations?, and (2) Can these unusual properties be harnessed in novel electronic materials? To address these questions, we first examine the distribution of $\pi$ electrons in the various porphyrin nanorings. Figure 3 shows a schematic of individual nanorings containing a variety of porphyrin units, and the $\pi$ electrons that form a continuous conjugated pathway within these nanorings are highlighted for clarity. The highlighted conjugation pathways are consistent with a previous study on other porphyrin-based systems by Sessler et al.[26] and are also consistent with Scott's definition of conjugated belts that are "distinguished by the presence of upper and lower edges that are conjugated but never coincide."[27] As shown in Figure 3, each porphyrin subunit contributes 14 $\pi$ electrons within the conjugated pathway; specifically, nanorings composed of 1, 2, and 3 porphyrin units contain 14, 28, and 42 $\pi$ electrons, respectively. Most importantly, nanorings containing an odd number of porphyrin units are aromatic and satisfy Hückel's (4*n* + 2) π-electron rule (where *n* is an integer), whereas nanorings composed of an even number of porphyrin units are non-aromatic and follow the (4*n*) π-electron rule instead.

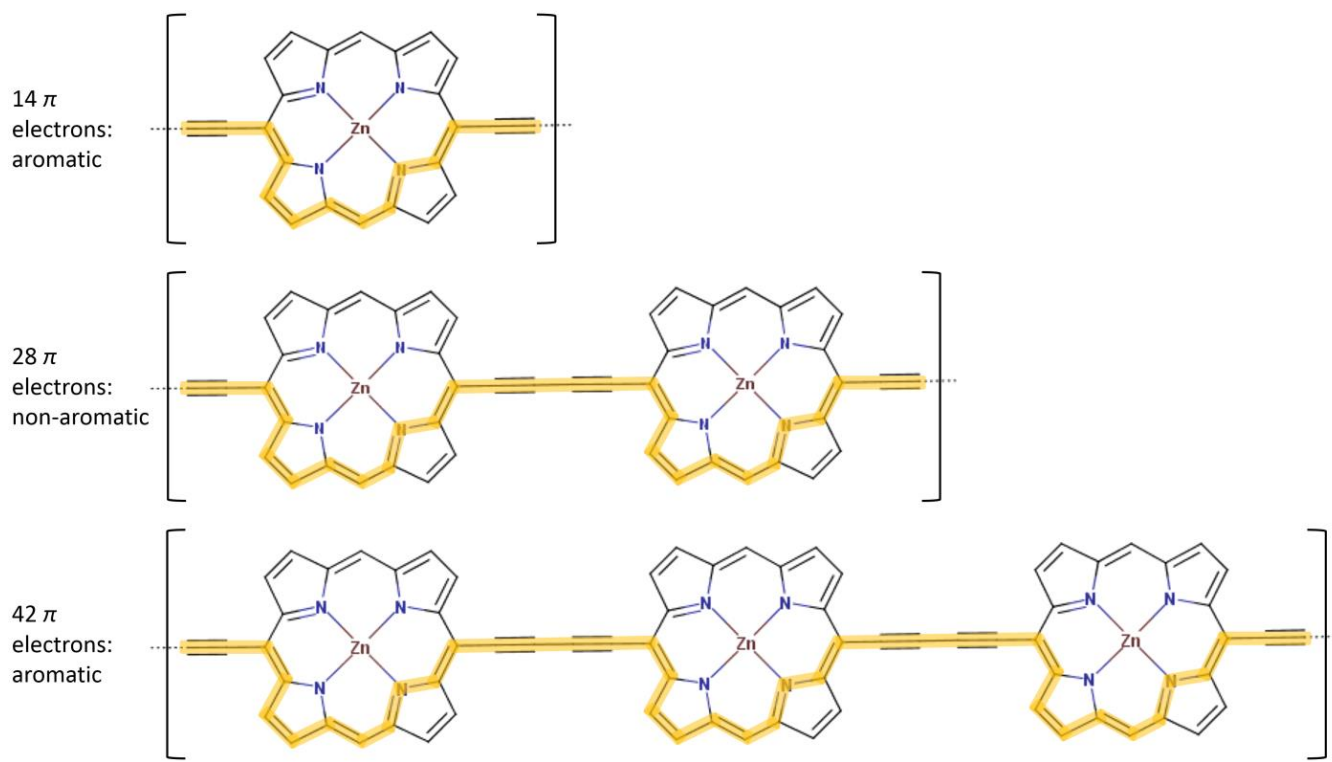

**Figure 3.** Schematic of nanorings composed of 1, 2, and 3 porphyrin units. The π electrons that form a continuous conjugated pathway within these nanorings are highlighted for clarity in orange – nanorings containing an odd number of porphyrin units are aromatic and satisfy the (4*n* + 2) π-electron Hückel rule, whereas nanorings composed of an even number of porphyrin units follow the non-aromatic (4*n*) π-electron rule.

By definition, aromatic systems are electronically more stable compared to their non-aromatic counterparts, and these results strongly imply that nanotubes containing an odd number of porphyrin units are more stable than their even-number counterparts. Indeed, our calculations corroborate these effects, and Figure 4 plots the energy of the valence band maximum for the porphyrin nanotubes as a function of size. At zero Kelvin, the chemical potential (i.e., the Fermi energy) lies just above the valence band maximum and, therefore, the VBM energies are a direct measure of the relative electronic stabilities of the various porphyrin nanotubes. As shown in Figure 4, nanotubes containing an odd number of porphyrin units are indeed more stable (i.e., more negative) than their even-numbered counterparts, which is consistent with the aromatic/non-aromatic properties of even/odd-numbered nanostructures discussed previously.

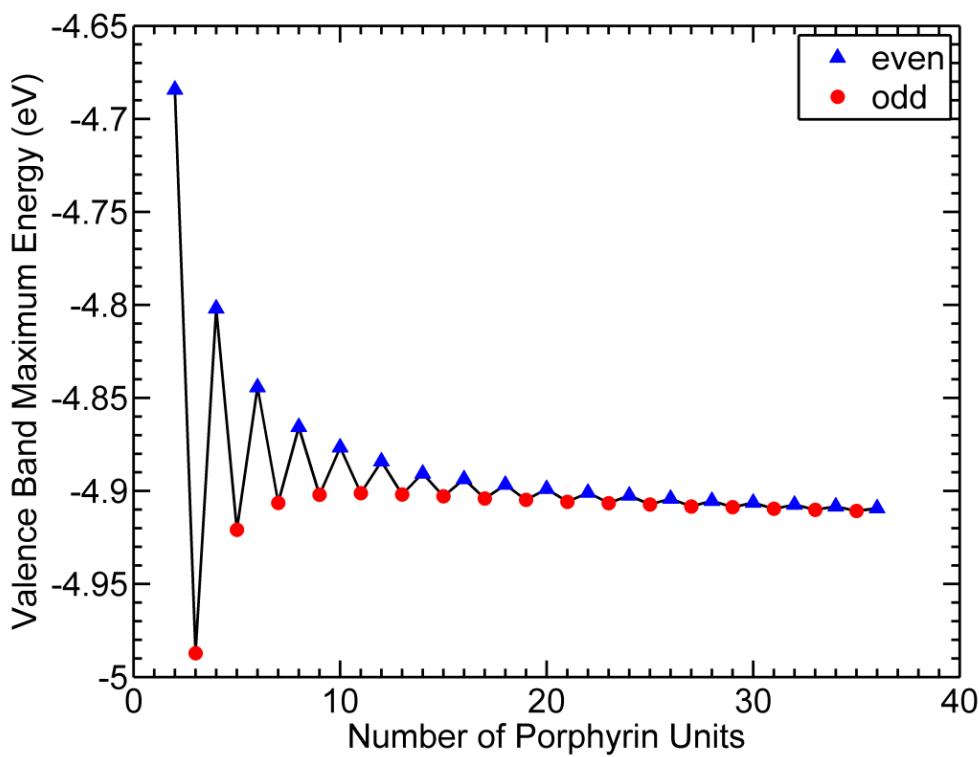

**Figure 4.** Plot of the valence band maximum energy as a function of the number of porphyrin units.

To give further insight into whether these unusual aromatic/non-aromatic effects can be experimentally observed, we now examine the electronic band structures and orbital diagrams in these porphyrin nanotubes. Figure 5 plots the electronic band structures of selected porphyrin nanotubes along the irreducible Brillouin zone defined by the high-symmetry points $\Gamma$ and X in momentum space (band structures and effective hole masses for all porphyrin nanotubes P02 – P36 can be found in the Supporting Information). In all of these nanostructures, we find that the electronic band structures are characterized by a direct band gap at the X point with a nearly flat

(dispersionless) conduction band. This conduction state originates from the dispersionless band between the X and M symmetry points in the parent porphyrin sheet (see Supporting Information) and reflects extremely localized (i.e., non-conducting) orbitals in the conduction band.

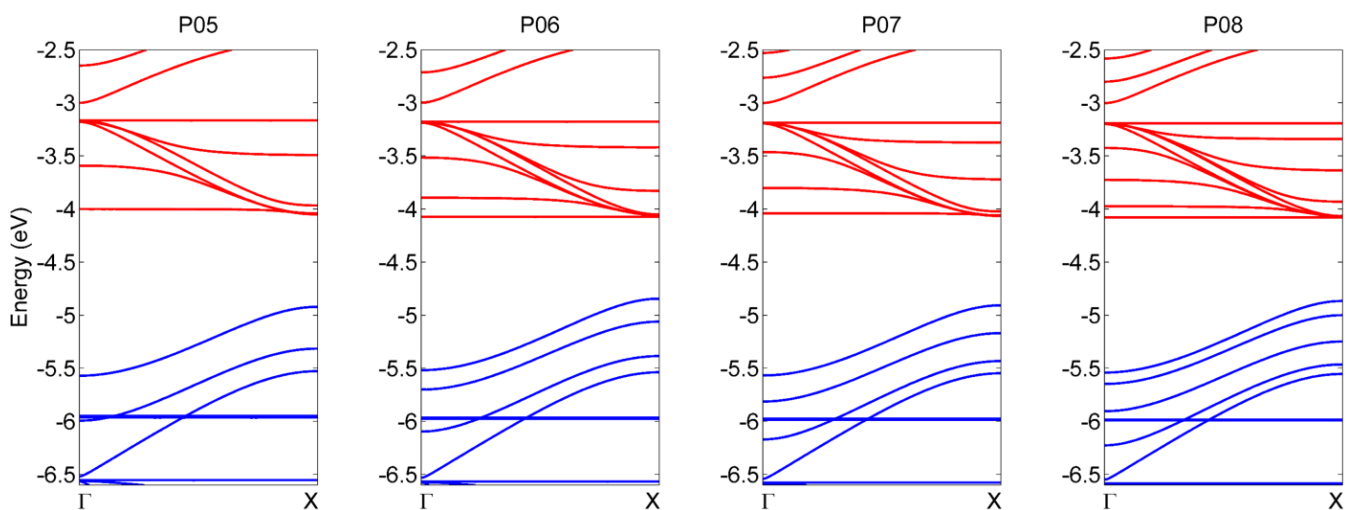

**Figure 5.** Electronic band structures (relative to vacuum at 0 eV) of selected porphyrin nanotubes. All of the nanotubes possess a direct band gap at the X point and a nearly dispersionless conduction band. Even within the plotted energy range, the oscillations of the valence band maximum energy between P05 – P07 can be observed.

Figure 6 shows the highest occupied molecular orbital (HOMO) and lowest unoccupied molecular orbital (LUMO) at the X symmetry point for the P5 and P6 nanotubes. In both cases, the HOMO and LUMO have different electronic symmetries, indicating that electronic transitions from the HOMO to the LUMO are optically active "bright" states and could be experimentally observed (although other electronic effects besides symmetry may play additional experimental roles). In other words, the bandgap oscillations in these porphyrin nanotubes, which are intrinsically due to alternating aromatic/non-aromatic properties, are optically allowed and should be readily detectable in UV photoelectron spectroscopic experiments. It is important to point out that while the HOMOs for both nanotubes are localized on the staves parallel to the tube axis and around the circumference, the LUMOs are markedly different for each tube. Specifically in the P5 structure, the LUMO is localized on the staves parallel to the tube axis, while the LUMO in P6 is localized on the staves around the circumference. As a result, nanotubes containing an odd number of porphyrin units can be optically excited by light polarized along the tube axis, whereas nanotubes with an even number of porphyrin units can only be excited by light polarized (or circularly polarized) perpendicular to the tube axis.

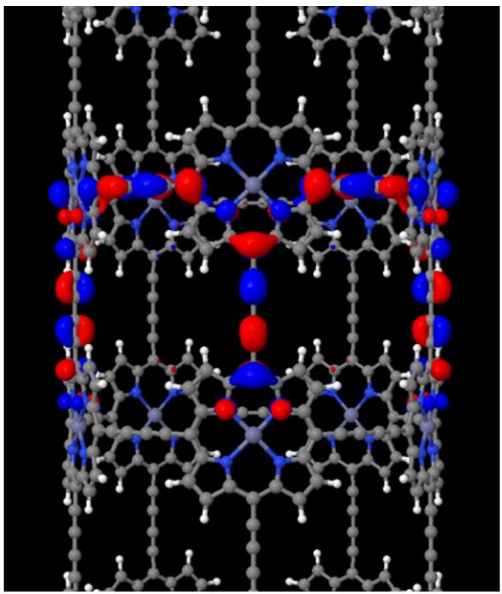
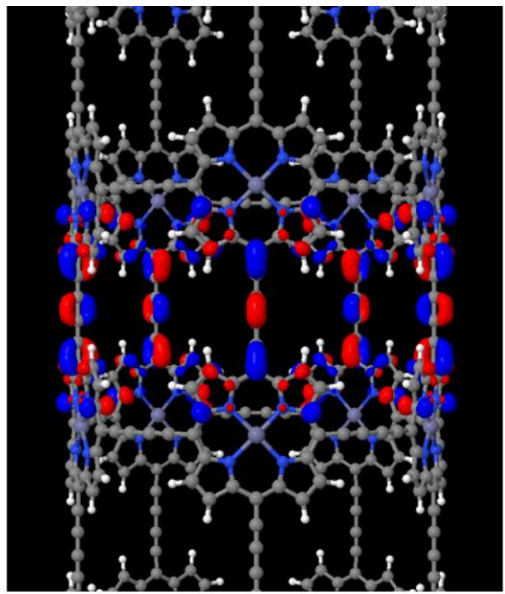

P5 HOMO    P5 LUMO

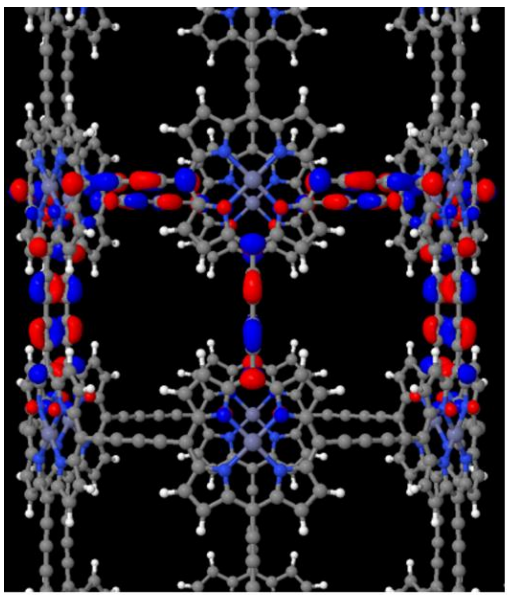
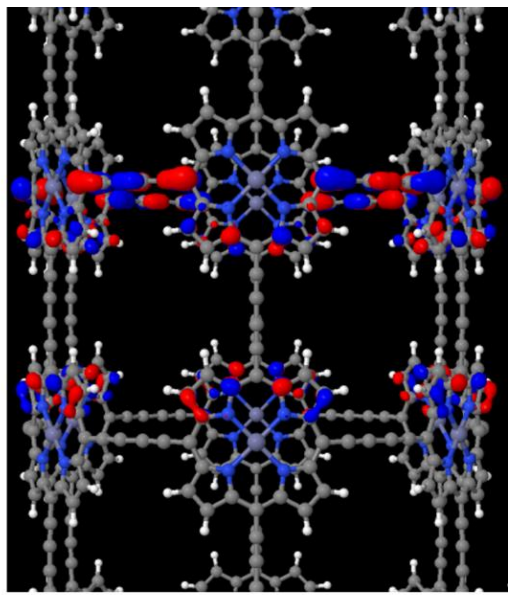

P6 HOMO    P6 LUMO

**Figure 6.** Plot of the highest-occupied and lowest-unoccupied orbitals at the X symmetry point for the P5 and P6 nanotubes (only orbitals within one unit cell are shown for clarity). While the HOMOs for both structures have similar orbital symmetries, the LUMOs are markedly different.

With these results and analyses, we are now finally in a position to address our previous question regarding whether the unusual properties of these nanotubes can be harnessed in novel electronic materials. In Figure 7, we plot the HOMO and LUMO energies of the individual monomers relative to vacuum for a selected pair of consecutive odd-numbered (3 and 5) and even-numbered (2 and 4) porphyrin nanotubes. In each of these p-n heterojunctions, upon photoexcitation, an exciton is created in which an electron is promoted into the LUMO. However, in the type I heterojunction formed by odd-numbered porphyrin units, both the electron and hole migrate toward the acceptor, resulting in charge recombination and light emission. Conversely in the type II heterojunction formed by even-numbered porphyrin units, photoexcitation will result in electron transfer to the LUMO acceptor and simultaneous hole-transfer to the HOMO of the donor. In this situation, the electrons and holes can be separated and driven towards different electrodes to avoid charge recombination. It is important to mention that while Figure 7 depicts heterojunctions with consecutive odd- and even-numbered units (i.e. P3/P5 and P2/P4, respectively), heterojunctions with *even larger* energy differences can be constructed using different combinations of nanotube sizes. For example, we obtain a conduction band offset of $\Delta E = 0.05$ eV for a heterojunction composed of P3 and P15). Furthermore, to confirm that these energy differences are indeed characteristic of these nanotubes and are not due to numerical error (or even artifacts of the exchange-correlation functional used), we also performed calculations with the PBE functional and found similar trends and a nearly identical conduction band offset of $\Delta E = 0.02$ eV for the P3/P5 heterojunction (see Supporting Information). The most salient point of these results is that *both* of these p-n heterojunctions (type I *and* type II) are possible and present in the porphyrin nanotube family, whereas only the type I heterojunction is present in conventional CNTs within the same chirality family. More importantly, since the porphyrin nanotubes are synthesized from a bottom-up, Venier template-directed approach, these nanotubes allow for a *controlled* synthesis of donor-acceptor, p-n heterojuctions with the desired energy alignments and properties (a situation that is not possible with conventional CNTs currently obtained with a random distribution of chiralities and electronic properties).

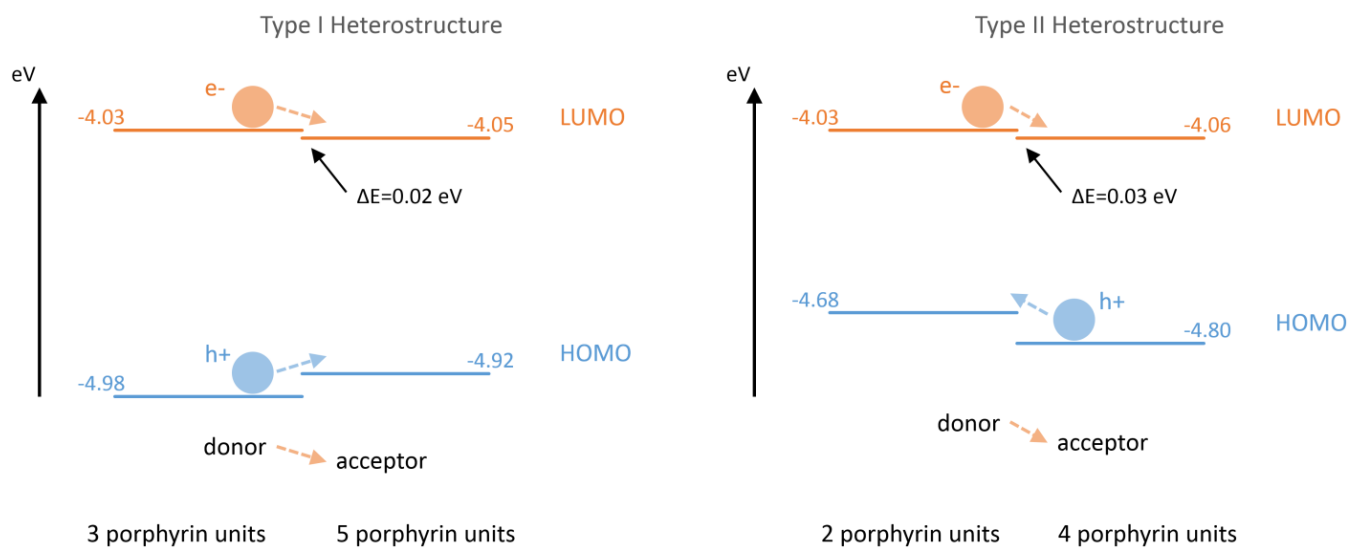

**Figure 7.** HOMO–LUMO energy level alignments found in consecutive odd-numbered (3 and 5) and even-numbered (2 and 4) porphyrin nanotubes. Upon photoexcitation, both the electron and hole migrate towards the acceptor and re-combine in the type I heterostructure (left) whereas the electron and hole can be separated in the type II heterostructure (right).

In conclusion, we find that the electronic properties of these template-directed porphyrin nanotubes exhibit unusual effects, and our calculations have several implications for experiments. First using large-scale DFT calculations, we surprisingly observe *extremely* large oscillations (spanning a range up to 0.32 eV) in the electronic bandgap of these nanostructures as a function of size. These oscillations are in contradiction to typical quantum confinement effects but are large enough to be readily detectable in UV photoelectron spectroscopic experiments. Second, we find that these intriguing electronic oscillations arise from a size-dependent alternation of aromatic/non-aromatic characteristics in these porphyrin nanotubes – a unique property that is not present in conventional CNTs. Third, in all of these nanotubes, electronic transitions from the HOMO to the LUMO are direct-bandgap, optically active "bright" states and can be experimentally observed (in contrast to cycloparaphenylene nanorings that would require an applied magnetic field to brighten the dark states via the Aharanov-Bohm effect[28]). Finally, we show that pairing odd- and even-numbered porphyrin nanotubes can give *both* type I and type II donor-acceptor p-n heterojunctions, a situation not found in conventional CNTs within the same chirality family. As such the "bottom-up" synthesis of these porphyrin nanotubes allows for a *controlled* fabrication of donor-acceptor, p-n heterojunctions with the desired properties that would have direct impact in electronic and photonic devices, such as nanoscale transistors, light-emitting diodes, and solar cells.

**Computational Details**

All calculations were carried out with a massively-parallelized version of the CRYSTAL14 program[29], which has the capability of using both all-electron Gaussian-type orbitals and exact Hartree-Fock exchange within periodic boundary conditions. The latter is particularly important for obtaining accurate electronic properties for periodic systems since the incorporation of Hartree-Fock exchange can partially correct for electron-delocalization errors inherent to both LDA (local density approximation) and GGA (generalized gradient approximation) exchange-correlation functionals. For this reason, we utilized the range-separated HSE06 functional[30] for obtaining the electronic properties for both the porphyrin sheet (see Supporting Information) and all of the porphyrin-based nanotubes. Most importantly, the screened HSE06 is more computationally efficient than conventional global hybrid functionals and is significantly more accurate than conventional semi-local functionals. It is worth noting that although the HSE06 calculations are more efficient than conventional hybrid DFT methods, the calculations on some of the largest porphyrin nanotubes were still extremely computationally intensive due to the immense size of these nanotubes. For example, the largest of these structures (specifically P36), consists of 1,476 atoms and 18,432 basis functions and, as such, this study constitutes the largest systematic study of these nanostructures to date. Geometries for all of the porphyrin nanotubes were optimized using a VDZ all-electron basis set with one-dimensional periodic boundary conditions along the tube axis. At the optimized geometries, a final single-point HSE06 calculation was performed with 100 k points along the one-dimensional Brillouin zone to obtain the electronic band structure for all of the nanotube geometries.

**Supporting Information**

The Supporting Information is available free of charge on the ACS Publications website at DOI: ____. Structural and electronic properties of the parent porphyrin sheet, structural properties for all porphyrin nanotubes: radii and relaxation energy, electronic properties for all porphyrin nanotubes: effective hole mass, bandgaps, and plots of electronic band structures, HOMO-LUMO energy level alignments for the P3/P5 heterojunction obtained with the PBE functional.

## Acknowledgements

We acknowledge the National Science Foundation for the use of supercomputing resources through the Extreme Science and Engineering Discovery Environment (XSEDE), Project No. TG-CHE150040.